\documentclass[twocolumn,showpacs,preprintnumbers,amsmath,amssymb]{revtex4}

\usepackage{graphicx}
\usepackage{dcolumn}
\usepackage{bm}
\newcommand {\apgt} {\ {\raise-.5ex\hbox{$\buildrel>\over\sim$}}\ }
\newcommand {\aplt} {\ {\raise-.5ex\hbox{$\buildrel<\over\sim$}}\ }

\begin{document}
 
  

 
\title{Critical exponents for the cloud-crystal phase transition \\
of charged particles in a Paul Trap} 

\author{D. K. Weiss, Y. S. Nam, and R. Bl\"umel}  
\affiliation{Department of Physics, Wesleyan University, 
Middletown, Connecticut 06459-0155}
  
\date{\today}

\begin{abstract} 
It is well known that 
charged particles stored in a Paul trap, one of the most versatile 
tools in atomic and molecular 
physics, may undergo a phase transition from a disordered cloud state 
to a geometrically well-ordered crystalline state (the Wigner crystal). 
In this paper we show 
that the average lifetime 
$\bar\tau_m$ of the metastable cloud state preceding 
the cloud $\rightarrow$ crystal phase transition follows 
a powerlaw, 
$\bar\tau_m \sim (\gamma-\gamma_c)^{-\beta}$, 
$\gamma>\gamma_c$, where $\gamma_c$ is the critical 
value of the damping constant $\gamma$ at which 
the cloud $\rightarrow$ crystal phase transition 
occurs. 
The critical exponent $\beta$ depends on the trap control 
parameter $q$, but is independent of the number of 
particles $N$ stored in the trap and 
the trap 
control parameter $a$, which determines the shape 
(oblate, prolate, or spherical) of the cloud. 
For 
%
$q=0.15,0.20$, and $0.25$, we find 
$\beta=1.20\pm 0.03$, 
$\beta=1.61\pm 0.09$, and
$\beta=2.38\pm 0.12$, respectively. 
In addition we find that for given $a$ and $q$, the critical 
value $\gamma_c$ of the damping scales approximately like 
$\gamma_c=C \ln [ \ln (N)] + D$ as a function of $N$, 
where $C$ and $D$ are constants. 
Beyond their relevance for Wigner crystallization 
of nonneutral plasmas in 
Paul traps and mini storage rings, we conjecture that 
our results are also of relevance for the field of 
crystalline beams.  
\end{abstract}



\maketitle

The Paul trap \cite{Paul1,PKG}  
is an electrodynamic device for storing charged particles 
for very long periods of time, free from contact with 
material walls. 
Trapping is achieved by applying suitable 
dc and 
ac voltages to the hyperbolic 
electrodes of the trap.
The resulting electric potentials create an 
effective potential minimum at the center of the 
trap, an immaterial trough that confines charged 
particles, in principle, forever. Storage times ranging 
from a few hours \cite{Nature} to a few days \cite{WSL} 
have been reported. 
 
Theoretically and experimentally, trapping of a 
single charged particle in an ideal Paul trap 
is understood 
in detail \cite{Paul1,PKG}, 
and even its quantum regime 
has already been explored  \cite{Brown,Wineland1}. 
However, if multiple 
particles are stored in the trap 
simultaneously, the Coulomb 
interactions between the particles 
cause their motions 
to be chaotic \cite{PKG,Hoff1,Nature}. 
In this case it is no longer 
possible to solve their equations of motion analytically. 
The chaotic motion of the particles has two consequences. 
(i) Due to the resulting high temperatures, we do 
not have to worry about quantum effects; a classical 
description of the trapped particles is sufficient. (ii) 
The chaotic motion of the particles in the trap 
causes the 
phenomenon of radio-frequency (rf) 
heating \cite{Nature,Kappler}. 
Damping must be imparted to this 
system to 
counteract the heating, whether through 
laser cooling \cite{Nature}, 
buffer gas cooling \cite{WSL}, 
or some other method, 
for instance cooling by the cold, neutral particles 
of a magneto-optic trap \cite{Win2}. With a 
relatively small damping, the rf heating power of the 
cloud will come into equilibrium with the cooling power, 
resulting from the damping mechanism, 
and a stationary-state gas cloud will result 
\cite{Nature,UNIV,MF}. 
However, with stronger damping, 
the heating of the cloud can be overcome, 
and the particles will transition into the  
crystalline phase \cite{Nature,WSL,MPQ1,NIST1}, 
the Wigner crystal \cite{Wigner}.  
We chose the Paul trap as 
a representative for a much wider class of periodically 
driven many-particle systems that also show 
Wigner crystallization and include 
particle accelerator beams \cite{cbeam1,cbeam2},
dusty plasmas \cite{dusty1,dusty2}, 
surface state electrons \cite{Andrei}, and 
colloidal suspensions \cite{Colloid1,Colloid2}. 
 
In this paper we use large-scale molecular dynamics simulations 
to show that the cloud $\rightarrow$ crystal phase transition may 
be interpreted as a critical phenomenon \cite{Baierlein} and we calculate the 
critical exponent. We also present a scaling law for the 
critical damping necessary to achieve crystallization in the Paul trap. 
 
The coupled equations of motion governing the 
dynamics of $N$ particles stored in the Paul trap, 
in dimensionless units \cite{UNIV}, are 
\begin{equation}
\begin{split}
\ddot{\vec r}_i + \gamma \dot{\vec r}_i + 
[a-2q\sin(2 \tau)]
 \left( \begin{matrix} x_i \\ y_i \\ -2 z_i \end{matrix} \right) \\
 =   \sum_{\substack{j=1\\j\neq i}}^N 
 \frac{\vec r_i - \vec r_j}{|\vec r_i - \vec r_j |^3}, \ \ \ i=1,\ldots,N,
\label{Eq1} 
\end{split}
\end{equation} 
where $\vec r=(x,y,z)$, 
$\tau$ is the dimensionless time, 
$\gamma$ is the damping constant, 
$N$ is the number of 
trapped particles, and $a,q$ 
are the trap's dimensionless control parameters \cite{UNIV}, 
proportional to the dc and ac voltages applied 
to the trap's electrodes, respectively. 
The conversion between time $\tau$ and the number $n$ of 
rf cycles is accomplished via $n=\tau/\pi$. 
For given values of 
$N$, $a$, $q$, and $\gamma$,
we solve (\ref{Eq1}) numerically with a standard fifth-order 
Runge-Kutta integrator \cite{NumRec}.  
Each of our simulations starts at $\tau=0$ 
with randomly chosen initial conditions 
drawn from the phase-space box 
$-10 < x,y,z < 10$, $-1 < v_x, v_y, v_z < 1$ with 
a uniform distribution. We checked that, because 
of the chaotic nature of the 
particle dynamics in the trap, all of our results 
are completely insensitive to both the particular 
choice of 
random distribution and the size of the box. 
To monitor the progress of our simulations, 
we plot  $\langle x^2(\tau_n=n\pi)\rangle$,  $n$ integer, where 
the angular brackets indicate an ensemble average  over 
all $N$ particles. 
 
%
\begin{figure}[t]
\centering
\includegraphics[scale=0.6,angle=0]{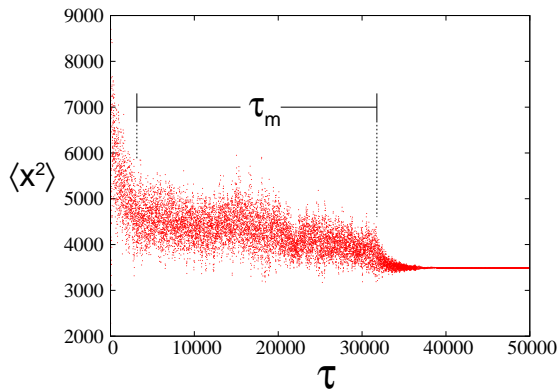}
\caption{\label{fig1} (Color online) Mean-square 
displacement $\langle x^2(\tau)\rangle$ of a 
100-particle cloud for 
$q=0.2$, $a=0.02$, and 
$\gamma=8.81\cdot 10^{-4} > \gamma_c=8.47\cdot 10^{-4}$. 
An initial transient (thermalization stage) is followed by a plateau of length 
$\tau_m$ (metastable state), which ultimately transitions 
into a state of constant  $\langle x^2(\tau)\rangle$
(flat line; crystalline state). 
}
\end{figure}
%
 
The result of a 
typical simulation run is shown in Fig.~\ref{fig1}. 
Since they are chosen at random, all of our 
initial conditions correspond to 
energetic particle clouds with large initial values 
of  $\langle x^2\rangle$ (see data points 
for $\tau\approx 0$ in Fig.~\ref{fig1}). 
However, because of the chaotic nature  
of its dynamics, the particle cloud very quickly 
loses the memory of its initial conditions and 
thermalizes.  
This corresponds to the  
initial transient [see the near-exponential 
decay phase over the first $\sim 1,000$ rf cycles 
($\tau\approx 0$ to $\tau\approx 3,000$) in 
Fig.~\ref{fig1}], followed by the 
establishment of a metastable stationary state 
(see the plateau in Fig.~\ref{fig1} of length 
$\tau_m\approx 28,000$), 
where the heating of the cloud comes into 
equilibrium with the damping. 
Following this, if, as in Fig.~\ref{fig1}, a 
relatively large $\gamma$ was chosen, 
the cloud eventually collapses into the crystal 
state. In Fig.~\ref{fig1} 
this final collapse manifests itself as the 
exponential decay phase immediately following the metastable 
state (to the right of the second dashed line in Fig.~\ref{fig1}) 
and ending in the crystalline phase, 
characterized by the  
absence of fluctuations in $\langle x^2\rangle$ for 
$\tau\gtrsim 40,000$. We checked explicitly that 
during its plateau phase the ion cloud is stable in 
the sense that there are no dynamical variables or 
expectation values that would decay during $\tau_m$. 
 
%
%
\begin{figure}
\centering
\includegraphics[scale=0.6,angle=0]{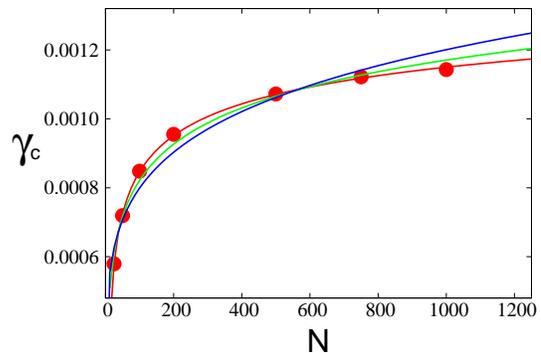}
\caption{\label{fig2} (Color online) Critical value $\gamma_c(N,q=0.2,a=0.02)$ 
of the damping 
constant $\gamma$ [see (\ref{Eq1})] as a function of $N$ at which the 
cloud $\rightarrow$ crystal phase transition occurs (red, closed circles).
The best-fitting powerlaw (blue, solid line), log-law 
(green, solid line), and the iterated-log-law (red, solid line) are also shown. 
Only the iterated-log-law, according to 
(\ref{GAMCLAW}), provides a satisfactory fit. 
}
\end{figure}
%

Confirming previous experimental 
\cite{Nature,WSL} and numerical \cite{ Nature} observations, 
we find that for given $N,a,q$ the 
cloud $\rightarrow$ crystal phase transition 
(the final collapse of the cloud in Fig.~\ref{fig1}) 
occurs in the vicinity of a critical value of $\gamma$, 
denoted by $\gamma_c$. In addition, 
corroborating earlier qualitative 
experimental observations 
(see, e.g., Fig.~3 in \cite{Nature}), 
we find that, for finite $N$, and a given finite 
simulation time $\tau_{\rm max}$, 
$\gamma_c$ is not sharply 
defined. Therefore, to determine $\gamma_c$ and its 
uncertainty, we proceed in the following way. 
For given $N,a,q$, we scan $\gamma$ from 
$\gamma_{\rm min} = 10^{-4}$ to 
$\gamma_{\rm max} =  2\cdot 10^{-3}$, a 
$\gamma$ interval that we know from experience 
contains $\gamma_c$ with certainty for $N$ ranging 
between 20 and 2,000 trapped particles. 
We find that in the interval  
$\gamma_{\rm min} < \gamma < \gamma_1(N,a,q)$, $N$-particle 
clouds are stable and never transition into 
the crystal. Following this is the interval 
$\gamma_1(N,a,q) < \gamma < \gamma_2(N,a,q)$, 
an interval of uncertainty, in which the clouds sometimes 
transition into the crystal and sometimes not. 
Adjacent to this is the interval 
$\gamma_2(N,a,q) < \gamma < \gamma_{\rm max}$, 
in which all $N$-particle clouds, independently of initial conditions, 
always transition into crystals. 
Defining 
$\Delta \gamma_c=\gamma_2-\gamma_1$ as the width of 
the uncertainty interval, we find that $\Delta\gamma_c$ 
shrinks, i.e., $\gamma_1$ and $\gamma_2$ both move toward each other, 
with increasing number $N$ of stored particles 
according to $\Delta\gamma_c \sim 1/\sqrt{N}$, 
and 
also with the maximal time $\tau_{\rm max}$ allowed for our 
simulations. To be practical, however, we limited the 
run time of our simulations to $\tau_{\rm max}=5\cdot 10^5 \pi$, 
very much larger than the typical decay time $1/\gamma$ 
of our system. We found that this choice of $\tau_{\rm max}$ 
yielded consistent results,
and we saw no need to increase $\tau_{\rm max}$. 
Having determined the uncertainty interval $[\gamma_1,\gamma_2]$, 
we define $\gamma_c=(\gamma_1+\gamma_2)/2$. 
 
As an example, for $q=0.2$, $a=q^2/2$, and $N$ ranging from 
25 to 1,000 particles, we plot, in Fig.~\ref{fig2}, 
the $\gamma_c$ values 
(red, closed circles) 
determined according to the numerical procedure described 
above. The uncertainty $\Delta\gamma_c$ of $\gamma_c$ 
is smaller than the size of the plot symbols in Fig.~\ref{fig2}. 
We found that 
neither a powerlaw ($\gamma_c=AN^B+C$, where 
$A,B,C$ are fit parameters;  blue, solid line in Fig.~\ref{fig2}) 
nor a log-law [$\gamma_c=A\ln(N)+B$, where $A,B$ are 
fit parameters; green, solid line in Fig.~\ref{fig2}] 
fits the $N$ dependence of 
$\gamma_c$ satisfactorily, but that the iterated-log-law 
\begin{equation}
\gamma_c(N,q=0.2,a=0.02)=C\ln[\ln(N)] + D 
\label{GAMCLAW}
\end{equation}
(red, solid line in Fig.~\ref{fig2}) provides an excellent fit, 
where $C=7.49\cdot 10^{-4}$ and $D=-2.97\cdot 10^{-4}$. 
For $N=100$, $\gamma_c=8.47\cdot 10^{-4}$. This is the reason 
for why the cloud in Fig.~\ref{fig1}, subjected to 
$\gamma=8.81\cdot 10^{-4} > \gamma_c$, ultimately collapses into 
the crystal state. 

At present, we are not able to provide an analytical 
explanation for the origin of the iterated-log scaling of $\gamma_c$.
However, the weak $N$-dependence of $\gamma_c$
may be understood qualitatively in the following way.
Since, in the large-$N$ limit, and 
close to the cloud $\rightarrow$ crystal phase transition point, 
charged particles in the interior of the Paul trap 
have a near-constant density (similar to a charged 
liquid in a confining harmonic-oscillator potential), 
all particles deep in the interior of the  trap may be treated 
as equivalent, since 
they are experiencing approximately the same homogeneous surrounding 
charge density. 
Given that $\gamma$ represents the energy loss per particle 
[see (\ref{Eq1})], 
$\gamma_c(N)$ is expected to be constant.
Thus, the small deviation of the $\gamma_c(N)$-scaling from constancy, 
i.e., the presence of the $\ln[\ln(N)]$ term, is a finite-size (surface) effect that is
hard to capture analytically. 
 
\begin{figure}
\centering
\includegraphics[scale=0.6,angle=0]{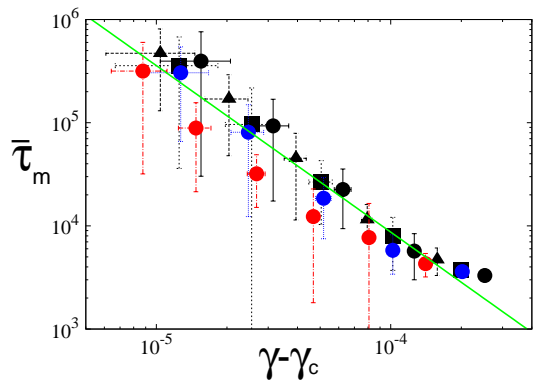}
\caption{\label{fig3} (Color online) Average time  $\bar\tau_m$ spent 
in the metastable state versus the
distance  $\gamma - \gamma_c$ from the critical point $\gamma_c$ for
$q = 0.2, a = 0$ (oblate), $a = q^2/2$ (spherical), and $a = 4q^2/5$
(prolate) (triangles, circles, and squares, respectively) and $N =
100, 200$, and 500 (black, blue, and red plot symbols, respectively).
The fit line corresponds to
$\bar\tau_m \sim (\gamma - \gamma_c)^{-1.61}$ 
       }
\end{figure}

 
We now turn to a more in-depth investigation of the 
cloud $\rightarrow$ crystal phase transition, i.e., 
we focus on the interval $\gamma > \gamma_2 > \gamma_c$. 
In particular, we are interested in the time it takes for a 
cloud to crystallize, once it has achieved its metastable 
state (the plateau in Fig.~\ref{fig1}), i.e., we are interested 
in the length of time $\tau_m$ the cloud spends in the 
metastable state before quickly transitioning into 
the crystalline state (ultimate exponential decay in 
Fig.~\ref{fig1}). It is intuitively clear that the larger 
$\gamma$, the shorter $\tau_m$. Conversely, 
when approaching $\gamma_2$ from above, and 
taking into account that clouds are stable for 
$\gamma < \gamma_1\approx \gamma_c$, 
$\tau_m$ should increase 
as $\gamma$ approaches $\gamma_2\approx \gamma_c$. 
According to the theory of 
critical phenomena \cite{Baierlein}, this suggests 
a powerlaw dependence according to 
\begin{equation}
\tau_m(N,a,q;\gamma) \sim [\gamma-\gamma_c(N,a,q)]^{-\beta(N,a,q)}  
\label{CREXP}
\end{equation}
for $\gamma \gtrsim \gamma_c$, 
where $\beta$ is the critical exponent. To find 
$\beta$ we ran our simulations for fixed $N,a,q$ for 
$\gamma$ values that approach $\gamma_c(N,a,q)$ from above 
and extracted $\tau_m$ via an automated, 
objective process \cite{auto}. 
Since the motion of the particles in the 
Paul trap is fully chaotic, 
small changes in the initial conditions can 
produce different values of $\tau_m$. Therefore, 
we ran our simulations 
with fifty different initial conditions 
and defined $\bar\tau_m$ as the average over the 
fifty resulting $\tau_m$ values. To characterize 
the statistical spread of the $\tau_m$ values, we 
also computed the standard deviation 
$\sigma = [(1/50)\sum_{j=1}^{50} (\tau_m^{(j)}-\bar\tau_m)^2]^{1/2}$. 
For $q=0.2$, Fig.~\ref{fig3} shows 
the resulting dependence of $\bar\tau_m$ on 
$(\gamma-\gamma_c)$ (plot symbols), where 
the error bars in Fig.~\ref{fig3} equal $\pm$ one standard 
deviation $\sigma$ for each corresponding data point. 
If (\ref{CREXP}) holds, the data 
in Fig.~\ref{fig3} should fall on a straight line. Within 
the error bars in Fig.~\ref{fig3} this is indeed the case 
and we extract $\beta=1.61\pm 0.09$ from Fig.~\ref{fig3}. 
We also notice that for the selected 
$q$ value, $\beta$ is approximately independent 
of $a$ and $N$.  

%
%
 
In order to investigate whether the 
observed powerlaw extends to other values of $q$, 
we ran additional simulations with $q=0.15$ and 
$q=0.25$.
For $q=0.15$ we obtained 
$\beta=1.20\pm 0.03$ and for 
$q=0.25$ we obtained 
$\beta=2.38\pm 0.12$. 
This supports the validity of the powerlaw
(\ref{CREXP}) with a critical exponent 
$\beta$ that depends only on $q$, 
but not on $a$ or $N$. According to the 
quoted uncertainties in $\beta$ we see that the 
critical exponent is more accurately defined for 
smaller values of $q$. The reason for this 
is straightforward. According to 
(\ref{Eq1}), $q$ determines the 
strength of the ac drive, which, in turn 
determines the degree of chaos in the trap. 
Therefore, smaller $q$ means less chaos, which 
implies smaller $\Delta \gamma_c$, which, in 
turn, results in a better defined $\beta$. 
  
%
%
%
%
 
But what about crystal $\rightarrow$ cloud transitions? 
Indeed, these were reported to occur 
as a function of reduced damping already 
in the earliest experiment on phase transitions in
a Paul trap \cite{WSL}. They are, however, 
of a completely different 
nature than the critical phenomena studied in this paper. 
Corroborating earlier results \cite{Nature,Kappler}, 
we found that in an ideal Paul trap described by
(\ref{Eq1}), even in the absence of damping (i.e. $\gamma=0$), 
crystal $\rightarrow$ cloud phase transitions do not
occur. We checked this fact explicitly for many 
different $a,q$ combinations, and $N$ ranging 
from 25 to 200. The explanation is straightforward. 
There is no chaos in the crystal state. Therefore, 
crystals do not heat, and are therefore stable 
even in the absence of damping. In experiments 
that do observe 
crystal $\rightarrow$ cloud transitions, the 
crystals are heated by an outside source, for 
instance by coupling to the hot, ambient air 
in the experiments reported in  \cite{WSL}. 
Thus, while 
crystal $\rightarrow$ cloud transitions certainly 
occur in experiments in which the crystals 
are coupled to a heat bath, 
their underlying mechanism is completely different 
from the purely dynamical transitions studied 
in this paper. Nevertheless, these transitions provide 
an important and natural extension of the zero-temperature 
phenomena studied here to finite-temperature critical 
phenomena in periodically driven multi-particle systems, 
a promising field for future research. 
 
While our results and methods are broadly applicable to a host of 
many-particle systems in various areas of 
physics, which also show Wigner 
crystallization \cite{cbeam1,cbeam2,dusty1,dusty2,Andrei,Colloid1,Colloid2}, 
{\it crystalline beams} \cite{cbeam1,cbeam2} 
are of particular importance. 
While some particle accelerators 
focus on the {\it energy frontier} (see, e.g., the Large Hadron Collider 
at CERN, Geneva) other particle accelerators focus on 
the {\it intensity frontier} (see, e.g., the Main Injector at Fermilab, 
Chicago). One way to approach the intensity 
frontier is via crystalline beams \cite{cbeam1,cbeam2}, 
the ultimate quality particle beams 
with the best possible brilliance and intensity. 
It is well-known that in their rest frame 
the dynamics of the beam particles are described by 
equations very similar to (\ref{Eq1}), and that a 
phase transition, 
very similar to the one described in 
this paper, induced by laser or electron cooling, 
precedes the transition into the Wigner crystal 
\cite{PB}. This phase transition has not yet been 
analyzed in terms of critical exponents and is an 
obvious and important example for testing and verifying 
the universality 
of our predictions. We expect that because of their 
immediate applicability to crystalline beams, our results 
will be of interest to the accelerator community. 
We mention that Wigner crystallization 
has already been observed in a mini storage ring 
\cite{cbeam3}.


In conclusion, we showed that for charged particles 
stored in a Paul trap, a 
critical value, $\gamma_c(N,a,q)$, of the 
damping constant $\gamma$ exists 
at which a cloud $\rightarrow$ crystal phase transition occurs. 
We showed that 
$\gamma_c$ scales approximately 
like $\ln[\ln(N)]$ in the number $N$ of stored 
particles in the trap. In addition, 
we showed that the cloud $\rightarrow$ crystal phase transition 
at $\gamma_c$ 
may be interpreted as a critical phenomenon with 
a critical exponent $\beta$ 
that predicts, given $\gamma$, the mean lifetime $\bar \tau_m$ of 
the metastable cloud before crystallization. 
Going beyond our atomic physics example, 
we conjecture that 
all damped, periodically driven, chaotic systems show a 
phase transition from a disordered (cloud) into an ordered 
(crystalline) state, characterized by powerlaws akin to 
the case of ions stored in a Paul trap discussed in this paper. 
Thus, the results reported in this paper are a step toward 
a comprehensive theory of 
phase transitions in periodically driven many-particle systems. 


\begin{thebibliography}{99}
\bibitem{auto} We 
 compute a moving average of $\langle x^2 \rangle_{\Delta t}$
 (averaging over a time interval $[t-\Delta t,t+\Delta t]$), which yields
 very sharp transitions suitable for an automatic and 
 objective extraction of $\tau_m$ via a computer program. 
 The choice of the interval $\Delta t$, which is of the order of the temporal
 transition width of the cloud $\rightarrow$ crystal transition, is not critical.
\end{thebibliography}
\end{document}